\documentclass[journal=jpclcd,manuscript=letter]{achemso}

\usepackage[version=3]{mhchem} 
\usepackage[T1]{fontenc}       
\usepackage{amsmath}
\usepackage[dvipsnames]{xcolor}

\author{Morgane Vacher}
\email{morgane.vacher@kemi.uu.se}
\affiliation[Uppsala University]
{Department of Chemistry -- \AA ngstr\"om, The Theoretical Chemistry Programme, Uppsala University, Box 538, 751 21 Uppsala, Sweden}
\author{Pooria Farahani}
\affiliation[Universidade de S\~ao Paulo]
{Instituto de Qu\'imica, Departamento de Qu\'imica Fundamental, Universidade de S\~ao Paulo, C.P. 05508-000, S\~ao Paulo, Brazil}
\author{Alessio Valentini}
\affiliation[Universit\'e de Li\`ege]
{D\'epartement de Chimie, Universit\'e de Li\`ege, All\'ee du 6 Août, 11, 4000 Li\`ege, Belgium}
\author{Luis Manuel Frutos}
\affiliation[Universidad de Alcalá]
{Departamento de Química Física, Universidad de Alcalá, E-28871 Alcalá de Henares, Madrid, Spain}
\author{Hans O. Karlsson}
\author{Ignacio Fdez. Galván}
\author{Roland Lindh}
\email{roland.lindh@kemi.uu.se}
\affiliation[Uppsala University]
{Department of Chemistry -- \AA ngstr\"om, The Theoretical Chemistry Programme, Uppsala University, Box 538, 751 21 Uppsala, Sweden}

\title[]{How Do Methyl Groups Enhance the Triplet Chemiexcitation Yield of Dioxetane?}

\begin{document}

\begin{tocentry}
\includegraphics[scale=1.0]{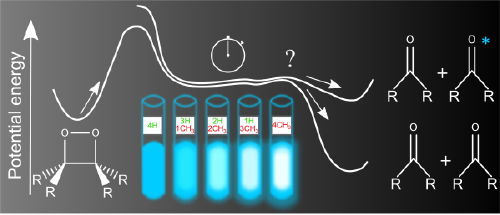}
\end{tocentry}

\begin{abstract}
Chemiluminescence is the emission of light as a result of a non-adiabatic chemical reaction. The present work is concerned with understanding the yield of chemiluminescence, in particular how it dramatically increases upon methylation of 1,2-dioxetane. Both ground-state and non-adiabatic dynamics (including singlet excited states) of the decomposition reaction of various methyl-substituted dioxetanes have been simulated. 
Methyl-substitution leads to a significant increase in the dissociation time scale. 
The rotation around the O-C-C-O dihedral angle is slowed down and thus, the molecular system stays longer in the ``entropic trap'' region. A simple kinetic model is proposed to explain how this leads to a higher chemiluminescence yield. 
These results have important implications for the design of efficient chemiluminescent systems in medical, environmental and industrial applications.
\end{abstract}


Rationalising the yields of chemical reactions in terms of simple and accessible concepts is one of the aims of theoretical chemistry. The formulation of such models is, however, challenging due to the complexity and high-dimensionality of the dynamics underlying chemical reactions. The present work is concerned with the yield of chemiluminescence, 
i.e. the emission of light as a result of a chemical reaction.~\cite{Matsumoto-2004} 
Today's basic understanding of chemiluminescence is that a thermally activated molecule reacts, and by doing so, undergoes a non-adiabatic transition~\cite{Yarkony-2012} to an electronic excited state of the product, which then releases the excess of energy in the form of light.
This fascinating phenomenon occurs in nature, in living organisms such as fireflies,~\cite{Navizet-2013} fungi~\cite{Purtov-2015} and fish~\cite{Widder-2010}; it is then called bioluminescence.~\cite{Navizet-2011} 
The emission of light has several uses: communication to attract partners, hunting by luring preys, defense to avoid predators, etc.~\cite{Widder-2010} 
Chemiluminescence is also a powerful tool in medicine, for instance for real-time in vivo imaging,~\cite{Daunert-2006} and in other fields for biosensing for environmental pollutants, food industry, etc.~\cite{Ripp-2003} A fundamental and outstanding challenge is to understand what determines the yield of chemi- and bio-luminescence, i.e. the amount of photons emitted per reacted molecules.

\begin{figure}
  \includegraphics[scale=0.25]{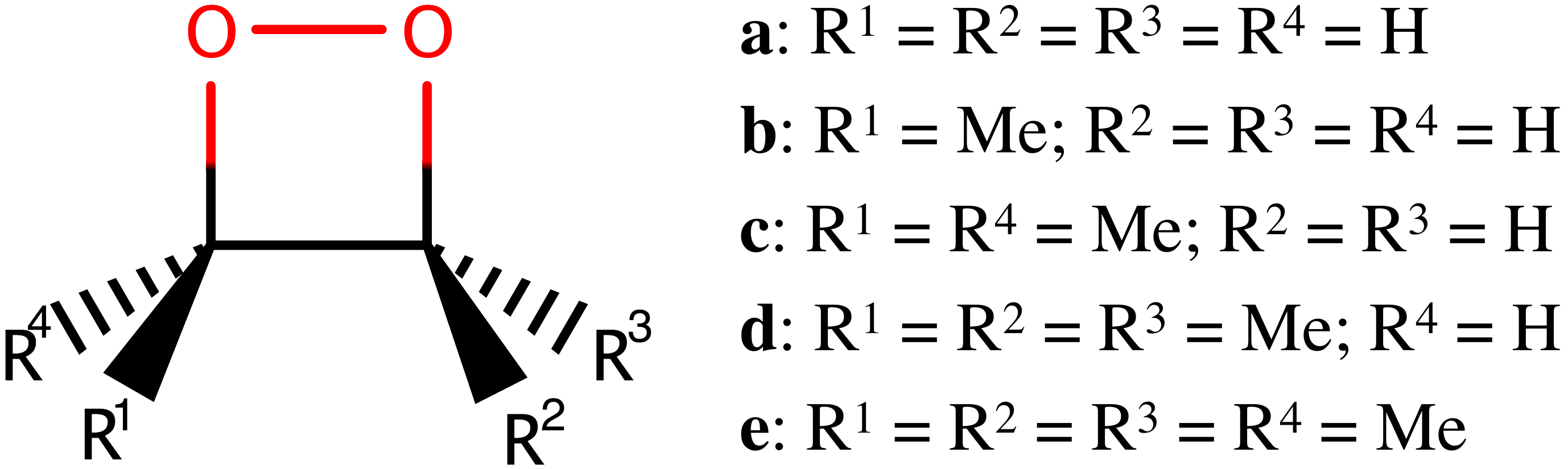}
  \caption{
  Dioxetane molecules where the hydrogen atoms are systematically substituted with methyl groups. 
  }
  \label{Fig_Mol}
\end{figure}

Almost all currently known chemiluminescent systems have the peroxide bond --O--O-- in common, the smallest being the 1,2-dioxetane molecule. The general mechanism of chemiluminescence in 1,2-dioxetane consists of two steps:~\cite{DeVico-2007,Farahani-2013,Vacher-2017-JCTC} (i) the O--O bond breaks leading to a biradical region where at least four singlet and four triplet states lie close in energy and (ii) the C--C bond breaks leading to dissociation into two formaldehyde molecules, which can end up in the electronic ground state, or in a singlet / triplet excited state. Already in the 1980's, dioxetane molecules with systematic substitution of a hydrogen atom by a methyl group (Figure~\ref{Fig_Mol}) were studied experimentally to try to rationalise chemiluminescence yields.~\cite{Adam-1985} Singlet and triplet excitation yields were determined by both chemiluminescence methods and chemical titration methods.
The experiments showed that the yield of the triplet excited states is much higher than that of the singlet excited states. 
An important result is that the excitation yield increases significantly with the degree of methylation: substituting all four hydrogen atoms by methyl groups enhances the chemiluminescence yield from approximately 0.3\% to 35\%.~\cite{Adam-1985} More than 30 years after the measurements, the reason for the impressive increase in chemiluminescence yield with the degree of methylation remains an outstanding question. The aim of the present work is to address it and explain it with concepts available to any chemist. This is an important question not only for understanding the chemiluminescence in the dioxetane molecules but also for understanding how Nature has designed such efficient bioluminescent systems as found in living organisms, and how mankind can design potentially more efficient chemical systems useful in medical applications for instance.

Born-Oppenheimer and non-adiabatic dynamics simulations of the (un-methylated) 1,2-dioxetane molecule~\cite{Vacher-2017-JCTC} have recently demonstrated 
that an ``entropic trap'' regulates the outcome of the dissociation, by delaying the exothermic ground state dissociation and by giving the molecule time to access excited states for instance. 
It was suggested in a previous theoretical study~\cite{DeVico-2007} that the addition of substituents would increase the time spent in the entropic trap through the increase of the number of degrees of freedom; this would then enhance the possibility to populate the product excited states. 
With simulations of the actual dynamics of five of the different dioxetane molecules \textbf{a}-\textbf{e}, the present work demonstrates that dissociation does take longer time upon methylation. However, this is partly due to a simple mass effect. A kinetic model is also presented to explain how slower dissociation can lead to a higher chemiluminescence yield.



To allow comparison between the different chemical compounds, the dynamics is initiated and simulated the same way for all.
The approach used is the same as in the recent work on the (un-methylated) 1,2-dioxetane molecule:~\cite{Vacher-2017-JCTC} the trajectories are initialised and propagated from the transition state (TS) for the O--O bond breaking (since it controls the overall reaction rate),  
by giving a small amount of kinetic energy (1~kcal/mol) along the reaction coordinate towards the biradical region.~\cite{Lourderaj-2008,Sun-2012,Farahani-2013,Vacher-2017-JCTC,Newton-X} 
Positions and momenta along all normal modes (other than the reaction coordinate) are sampled from a Wigner distribution, 
using the Newton-X package~\cite{Newton-X}. 
Born--Oppenheimer dynamics and non-adiabatic surface hopping dynamics (including transitions among all the four lowest-energy singlet states with the Tully's fewest switches algorithm~\cite{Tully-1990}) are simulated with a time step of 10~a.u. ($\approx$ 0.24~fs).
The decoherence correction proposed by Granucci and Persico is used with a decay factor of 0.1~hartree.~\cite{Granucci-2007} The implementation of the above methods in a development version of the Molcas package is used.~\cite{Molcas-2016}
For all compounds, all nuclear coordinates are taken into account; it amounts to 24 nuclear Cartesian coordinates for the un-methylated 1,2-dioxetane and 60 for the tetra-methylated 1,2-dioxetane. 
The electronic structure method used 
is the complete active space self-consistent field (CASSCF)~\cite{Roos-1980} method state-averaging over the four lowest-energy singlet states equally. The active space used consists of 12~electrons and 10~orbitals: the four $\sigma$ and four $\sigma^*$ orbitals of the four-membered ring, plus the two oxygen lone-pair orbitals perpendicular to the ring. The ANO-RCC basis set with polarised triple-zeta contraction
~\cite{Roos-2004} and the atomic compact Cholesky decomposition (acCD)~\cite{Aquilante-2009} auxiliary basis sets (for representing the two-electron repulsion integrals) are used.

\begin{figure}
  \includegraphics[scale=1]{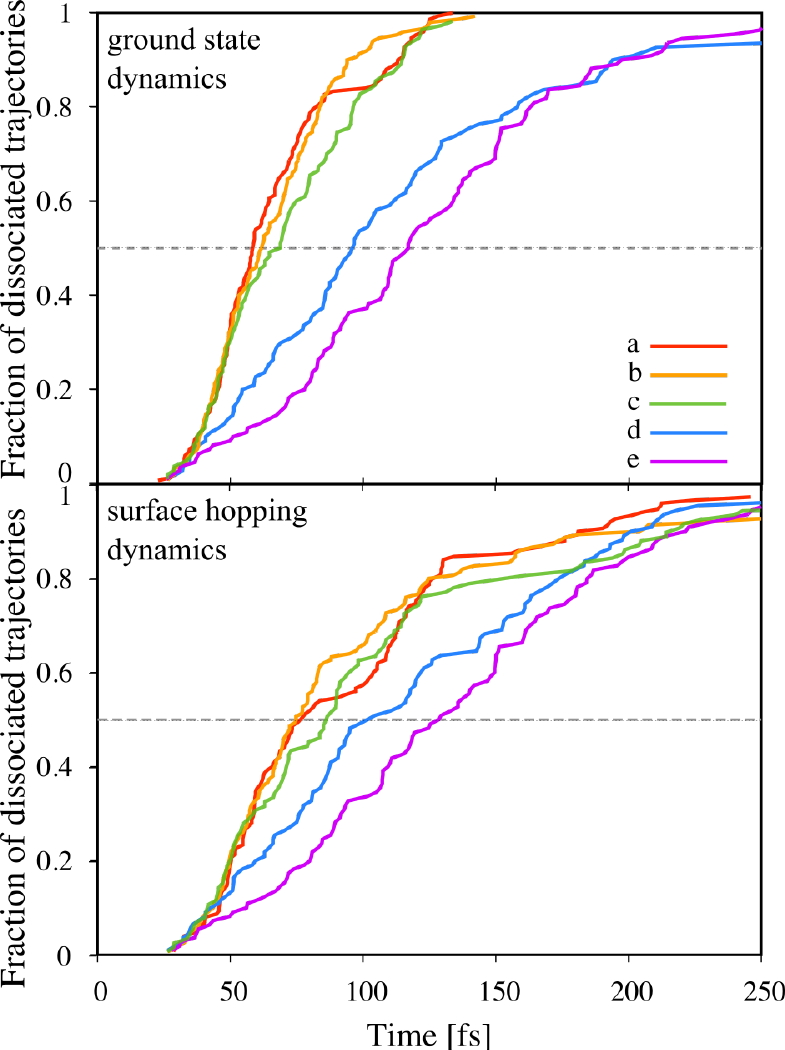}
  \caption{Dissociation time scale of an ensemble of 110 ground-state trajectories (upper panel) and surface-hopping trajectories including the four lowest-energy singlet states (lower panel), for the compounds   \textbf{a} (red),  \textbf{b} (yellow), \textbf{c} (green),  \textbf{d} (blue) and \textbf{e} (purple). The horizontal dashed lines indicate dissociation of half of the trajectories.}
  \label{Fig_Dissoc}
\end{figure}

First, the results of Born-Oppenheimer ground state dynamics are presented. Out of the three isomers with two methyl groups, only compound \textbf{c} (in which the methyl groups are attached to the same carbon) is studied. For each of the five studied compounds, an ensemble of 110 trajectories was run. It is noted that 
none of the trajectories initially directed towards the product recrossed the TS for any of the compounds. 
Figure~\ref{Fig_Dissoc} (upper panel) shows the time evolution of the fraction of ground state trajectories that have dissociated. 
Dissociation was considered to occur when the central C--C bond length exceeded 2.4 \AA~(two times the van der Waals radius of a carbon atom).
~\cite{Vacher-2017-JCTC} 
For all compounds, dissociation starts to occur from approximately $t=30$~fs. However, the subsequent dissociation dynamics time scale differs for the different compounds. In table~\ref{Tab_Dissoc}, the dissociation half-times, i.e. times required for half of the trajectories to have dissociated, are given. The general trend is that the dissociation half-time increases with the degree of methylation of 1,2-dioxetane; $t_{1/2}^{BO}=58.6$~fs for the un-methylated 1,2-dioxetane (compound \textbf{a}), while $t_{1/2}^{BO}=116.9$~fs for the tetramethylated 1,2-dioxetane (compound \textbf{e}).

\begin{table}
 \caption{Dissociation half-times [fs] from ground state and surface hopping dynamics simulations for the different compounds.}
  \label{Tab_Dissoc}
 \begin{tabular}{||c | c | c | c | c | c ||} 
 \hline
 Compound & \textbf{a} & \textbf{b} & \textbf{c} & \textbf{d} & \textbf{e} \\ 
 \hline\hline
 $t_{1/2}^{BO}$ & 58.6 & 62.7 & 69.2 & 95.8 & 116.9 \\ 
 \hline
 $t_{1/2}^{S}$ & 75.3 & 74.3 & 86.4 & 101.6 & 128.5 \\
 \hline
\end{tabular}
\end{table}

\begin{figure}
  \includegraphics[scale=1.0]{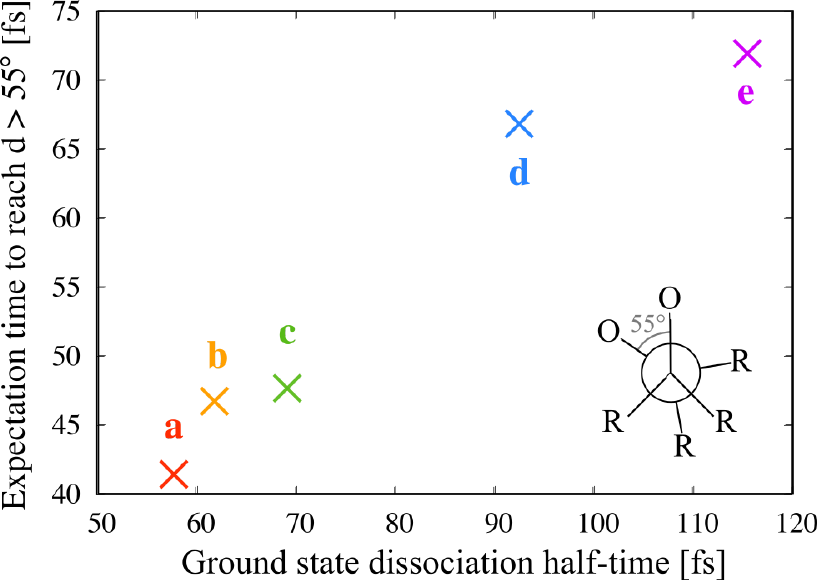}
  \caption{Time required for the O-C-C-O dihedral angle (averaged over the ensemble of 110 ground state trajectories) to reach a value greater than 55$^\circ$ (as shown in insert, bottom right) versus the ground state dissociation half-time of the ensemble of trajectories, for the compounds \textbf{a} (red),  \textbf{b} (yellow), \textbf{c} (green),  \textbf{d} (blue) and \textbf{e} (purple).}
  \label{Fig_Dih55}
\end{figure}

Why does dissociation take about twice longer for the tetramethylated 1,2-dioxetane than for the un-methylated 1,2-dioxetane? 
In a previous theoretical study~\cite{Vacher-2017-JCTC} of the decomposition of the un-methylated 1,2-dioxetane, it was demonstrated the existence of specific geometrical conditions for the trajectories to be able to escape from the entropic trap and for dissociation to be possible. In particular, it was shown that the O-C-C-O dihedral angle must be larger than approximately 55$^{\circ}$, otherwise the molecule remains trapped. Figure~\ref{Fig_Dih55} plots the time required for the O-C-C-O dihedral angle averaged over the ensemble of 110 ground state trajectories to exceed 55$^{\circ}$, for the different compounds. There is a clear correlation between this time and the dissociation half-time: the longer it takes for the O-C-C-O dihedral angle to exceed 55$^{\circ}$, the longer it takes to dissociate. In summary, upon methyl substitution, the torsional motion around the O-C-C-O dihedral angle is slower, the molecule stays longer trapped and therefore ground state dissociation occurs later.

\begin{figure}
  \includegraphics[scale=1]{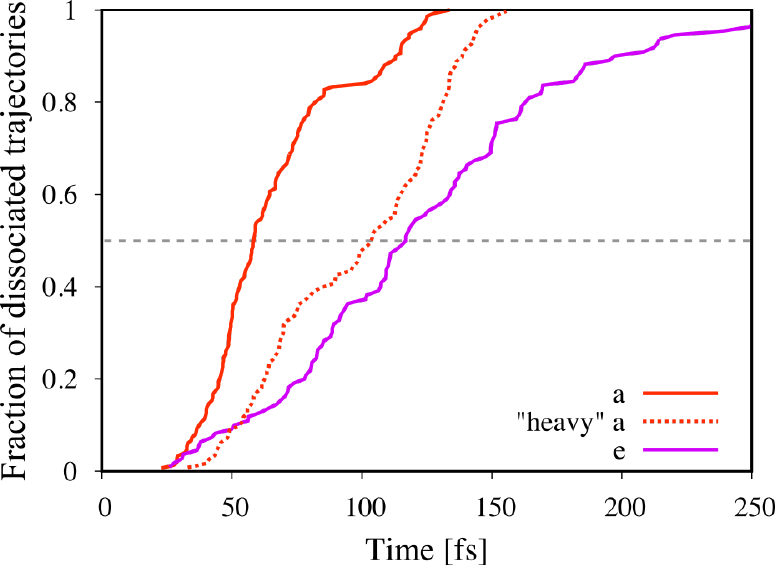}
  \caption{Dissociation time scale of an ensemble of 110 ground-state trajectories, for the compounds \textbf{a} (solid red), ``heavy'' \textbf{a} (dotted red) where the mass of the four hydrogen atoms was increased to reproduce the moment of inertia of the methyl groups CH$_3$, and \textbf{e} (purple). The horizontal dashed line indicates dissociation of half of the trajectories.}
  \label{Fig_Iso}
\end{figure}

Substituting an hydrogen atom by a methyl group can have several effects on the reaction dynamics. Is the longer trapping of the molecule due to the increase in the number of degrees of freedom, as suggested in a previous theoretical study?~\cite{DeVico-2007} Or is it due to heavier masses slowing down the motion? Or to steric effects? To investigate further the effect of methylation and to understand in particular the role of the mass, ground state dynamics is simulated for the un-methylated 1,2-dioxetane, but where the mass of the four hydrogen atoms was increased to 34.5 amu such as to reproduce the moment of inertia of the methyl groups CH$_3$. This way, the effect of the mass on the reaction dynamics is isolated from the other effects, such as steric effects. Figure~\ref{Fig_Iso} shows the time evolution of the fraction of ground state trajectories that have dissociated, for compounds \textbf{a}, ``heavy'' \textbf{a} (where the mass of the four hydrogen atoms was increased such as to reproduce the moment of inertia of the methyl groups CH$_3$) and \textbf{e}. 
The dissociation half-time for the ``heavy'' 1,2-dioxetane is $t_{1/2}^{BO}=102.8$~fs, almost as long as for the compound \textbf{e}. Dissociation occurs more slowly in compound \textbf{e} especially at longer times (after $t=150$~fs). 
The simulations suggests that approximately 75\% of the increase in the dissociation half-time of compound \textbf{e}, compared to compound \textbf{a}, is actually due to a pure mass effect. (This is in contrast to the hypothesis put forward in a previous theoretical study.~\cite{DeVico-2007}) In particular, heavier groups on the carbon atoms slow down the rotation around the O-C-C-O dihedral angle, and postpone the possible exit from the entropic trap. The rest of the increase in the dissociation time scale may be due to steric effects between the methyl groups for instance.

Now, the role of the three singlet excited states in the decomposition reaction is investigated. 
Figure~\ref{Fig_Dissoc} (lower panel) shows the time evolution of the fraction of surface hopping trajectories that have dissociated.
For all compounds, the singlet excited states postpone the dissociation further by 6 to 17~fs (table~\ref{Tab_Dissoc}). 
Over the five ensembles of 110 surface hopping trajectories, only one is observed to be dissociated on the singlet first excited states $S_1$. 
This is expected because of the extremely low singlet states excitation yield 
(0.25\% even for the decomposition of compound \textbf{e})~\cite{Adam-1985}. The present results are thus consistent with the experimental observations. It is noted that, based on ensembles of 110 trajectories, the dissociation half-time of compound \textbf{b} is calculated to be 1~fs shorter than that of compound \textbf{a}. 
Yet, importantly, the general trend which consists of an extension of the dark decomposition time scale with the degree of methylation, is preserved.

\begin{figure}
  \includegraphics[scale=0.97]{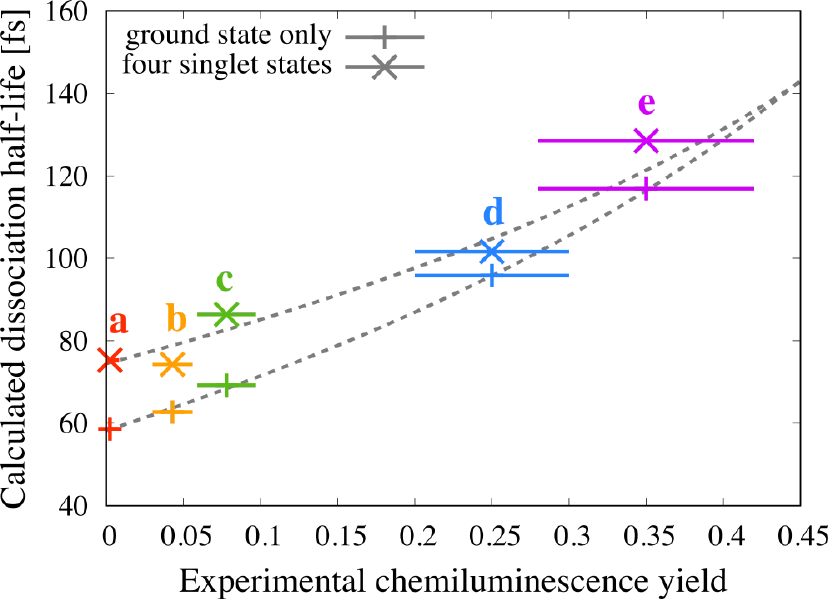}
  \caption{Simple kinetic model fitting the experimental~\cite{Adam-1985} triplet excitation yield $\phi$ and the calculated dissociation half-time $t_{1/2}$ using adiabatic ground state dynamics simulations (bar) or non-adiabatic surface hopping dynamics simulations (cross), for the compounds \textbf{a} (red),  \textbf{b} (yellow), \textbf{c} (green),  \textbf{d} (blue) and \textbf{e} (purple). The horizontal lines represent the experimental error bars.~\cite{Adam-1985} The dashed curves correspond to the analytical expression $t_{1/2}=t_0- \frac{\ln(1-4 \phi/3)}{4k}$ where the parameters $t_0$ and $k$ were fitted to the data, taking into account the experimental error bars: $t_0=58$ fs and $k=0.0027$ fs$^{-1}$ for ground state, and $t_0=74$ fs and $k=0.0033$ fs$^{-1}$ for surface hopping calculations.}
  \label{Fig_Model}
\end{figure}

In summary, heavier substituents on the carbon atoms slow down the torsional motion around the O-C-C-O dihedral angle, which traps the molecule for longer time and postpones ground state dissociation. But how does this lead to higher excitation and chemiluminescence yields? A simple kinetic model is proposed to explain how the entropic trap determines the chemiluminescence yield. 
In the following, $S$ represents the manifold of the four lowest-energy singlets and $T$ the manifold of the four lowest-energy triplets. The excitation yield to singlet states being two to three orders of magnitude lower than the excitation yield to triplet states,~\cite{Adam-1985} chemiluminescence is considered to be occurring through generation of $T$ only and the generation of singlet excited states is neglected. After breaking the O--O bond, the molecular system (initially in the singlet ground state $S_0$) enters the entropic trap region where $S$ and $T$ are degenerate. In this region, transfer of population between $S$ and $T$ occurs and eventually, at equilibrium (i.e. if the system stayed an infinitely long time in the entropic trap), the molecular system would distribute equally among the degenerate $S$ and three components of $T$. However, before reaching equilibrium, dark decomposition occurs on the ground state and interrupts the net transfer of population from $S$ to $T$ in the entropic trap region. 
The formation of products in triplet excited states and chemiluminescence yield are considered to be directly related to the population in $T$ in the biradical region, $x$, when ground state dissociation occurs.  
Assuming a first order kinetic model, with both forward and backward transfers of population possible and occurring with the same rate constant $k$, the population in $T$ (taking into account the three triplet components) after a time $t_{\text{trap}}$ spent in the entropic trap region reads:
\begin{equation}
x = \frac{3}{4} \left( 1- \exp(-4kt_{\text{trap}}) \right)
\end{equation}
Or, equivalently, the time needed to be spent in the entropic trap in order to populate $T$ with $x$ is:
\begin{equation}
t_{\text{trap}} = \frac{-\ln(1-4x/3)}{4k} \label{Eq_Model}
\end{equation}
The total dissociation time is the sum of $t_{\text{trap}}$ and $t_0$, the time necessary for the dissociation reaction to occur without spending any time in the entropic trap. The proposed explanation for the chemiluminescence yield is thus simply the following: the longer the system stays in the entropic trap, the more population is transferred from $S$ to $T$ and the higher the chemiluminescence yield is. In this simple model, the rate of population transfer $k$ is assumed to be the same for all compounds, i.e. unaffected by the methylation.
To test the model, both the calculated ground state and non-adiabatic surface hopping dissociation half-times are fitted to the experimental chemiluminescence yields using equation~(\ref{Eq_Model}). The results are shown in Figure~\ref{Fig_Model}. The simple kinetic model agrees quite nicely with the calculated and experimental data, supporting the interpretation of the results. Given the simplicity of the model, a better agreement is not expected.

To conclude: 
Ground state and non-adiabatic surface hopping dynamics simulations have shown that dark decomposition takes more time upon methylation of 1,2-dioxetane.
It was suggested before that this is due to the increase in the number of degrees of freedom. However, the simulations of the present work show that actually approximately 75\% of the increase is due to a simple mass effect. Heavier substituents on the carbon atoms slow down the nuclear motion, in particular the rotation around the O-C-C-O dihedral angle. A dihedral angle of at least 55$^{\circ}$ being necessary for escaping the entropic trap region, a longer time required to reach large dihedral angles means a slower dissociation. Simulations with ``frozen'' methyl groups would allow us to identify whether the rest of the increase in dissociation time scale is due to steric effects or more degrees of freedom. 
It is noted that only one trajectory over the five ensembles of 110 surface hopping trajectories was observed to dissociate on the singlet excited state. This is expected from the extremely low measured singlet excitation and fluorescence yields in 1,2-dioxetanes decomposition.~\cite{Adam-1985}

Slower dissociation means longer time spent in the entropic trap region, where the manifold of singlet states lies close in energy to the manifold of triplet states. This is where transfer of population between the singlet ground state and the triplet manifold occurs.  
Before reaching equilibrium among the degenerate electronic states, dark decomposition occurs on the ground state and interrupts the net transfer of population to the triplet states. The longer the system stays in the entropic trap, the more population is transferred from the singlet ground state to the triplet states and the higher the chemiluminescence yield is. A simple kinetic model has been proposed and tested by fitting the calculated dissociation half-times to the experimental chemiluminescence yields. It explains with accessible concepts the increase of the chemiluminescence yield upon methyl substitution. It is noted that our results are consistent with a previous experimental work where the chemiexcitation yield upon induced decomposition of substituted dioxetanes was observed to increase with the viscosity of the solvent.~\cite{Bastos-2013} There, the rotation around the O-C-C-O dihedral angle was suggested to compete with the electron back transfer necessary for such chemiexcitation.

The findings of the present work finally bring insights into chemiluminescence yields, and the substantial increase upon methylation. In particular, it demonstrates how substituents, through their mass simply, affect the dynamics of a reaction and as a consequence its yield. 
Future dynamics simulations including the population of the triplet states via spin-orbit coupling would allow for a more direct comparison between the simulated final populations in the triplet excited states and the experimental (not very low) triplet excitation yields. 

\begin{acknowledgement}
This work was supported by the Swedish Research Council (Grant 2016-03398). P.F. acknowledges the Fundaç\~ao de Amparo à Pesquisa do Estado de S\~ao Paulo (FAPESP) for the financial support under the 2015/02314-8 project number. L.M.F. acknowledges the Spanish MINECO (Grant CTQ2016-80600-P) for the financial support. The simulations were performed using resources provided by the Swedish National Infrastructure for Computing (SNIC) at UPPMAX and NSC centers.
\end{acknowledgement}

\begin{suppinfo}
Measured triplet excitation yields upon decomposition of methyl-substituted dioxetane molecules. Transition state structures and initial velocities along transition vectors. Derivation and assumptions of the kinetic model.
\end{suppinfo}

\bibliography{references}

\providecommand{\latin}[1]{#1}
\providecommand*\mcitethebibliography{\thebibliography}
\csname @ifundefined\endcsname{endmcitethebibliography}
  {\let\endmcitethebibliography\endthebibliography}{}
\begin{mcitethebibliography}{23}
\providecommand*\natexlab[1]{#1}
\providecommand*\mciteSetBstSublistMode[1]{}
\providecommand*\mciteSetBstMaxWidthForm[2]{}
\providecommand*\mciteBstWouldAddEndPuncttrue
  {\def\EndOfBibitem{\unskip.}}
\providecommand*\mciteBstWouldAddEndPunctfalse
  {\let\EndOfBibitem\relax}
\providecommand*\mciteSetBstMidEndSepPunct[3]{}
\providecommand*\mciteSetBstSublistLabelBeginEnd[3]{}
\providecommand*\EndOfBibitem{}
\mciteSetBstSublistMode{f}
\mciteSetBstMaxWidthForm{subitem}{(\alph{mcitesubitemcount})}
\mciteSetBstSublistLabelBeginEnd
  {\mcitemaxwidthsubitemform\space}
  {\relax}
  {\relax}

\bibitem[Matsumoto(2004)]{Matsumoto-2004}
Matsumoto,~M. Advanced chemistry of dioxetane-based chemiluminescent substrates
  originating from bioluminescence. \emph{J. Photochem. Photobiol. C:
  Photochem. Rev.} \textbf{2004}, \emph{5}, 27--53\relax
\mciteBstWouldAddEndPuncttrue
\mciteSetBstMidEndSepPunct{\mcitedefaultmidpunct}
{\mcitedefaultendpunct}{\mcitedefaultseppunct}\relax
\EndOfBibitem
\bibitem[Yarkony(2012)]{Yarkony-2012}
Yarkony,~D.~R. Nonadiabatic Quantum Chemistry---Past, Present, and Future.
  \emph{Chem. Rev.} \textbf{2012}, \emph{112}, 481--498\relax
\mciteBstWouldAddEndPuncttrue
\mciteSetBstMidEndSepPunct{\mcitedefaultmidpunct}
{\mcitedefaultendpunct}{\mcitedefaultseppunct}\relax
\EndOfBibitem
\bibitem[Navizet \latin{et~al.}(2013)Navizet, Roca-Sanju{\'a}n, Yue, Liu,
  Ferr{\'e}, and Lindh]{Navizet-2013}
Navizet,~I.; Roca-Sanju{\'a}n,~D.; Yue,~L.; Liu,~Y.-J.; Ferr{\'e},~N.;
  Lindh,~R. Are the Bio- and Chemiluminescence States of the Firefly
  Oxyluciferin the Same as the Fluorescence State? \emph{Photochem. Photobiol.}
  \textbf{2013}, \emph{89}, 319--325\relax
\mciteBstWouldAddEndPuncttrue
\mciteSetBstMidEndSepPunct{\mcitedefaultmidpunct}
{\mcitedefaultendpunct}{\mcitedefaultseppunct}\relax
\EndOfBibitem
\bibitem[Purtov \latin{et~al.}(2015)Purtov, Petushkov, Baranov, Mineev,
  Rodionova, Kaskova, Tsarkova, Petunin, Bondar, Rodicheva, Medvedeva, Oba,
  Oba, Arseniev, Lukyanov, Gitelson, and Yampolsky]{Purtov-2015}
Purtov,~K.~V.; Petushkov,~V.~N.; Baranov,~M.~S.; Mineev,~K.~S.;
  Rodionova,~N.~S.; Kaskova,~Z.~M.; Tsarkova,~A.~S.; Petunin,~A.~I.;
  Bondar,~V.~S.; Rodicheva,~E.~K. \latin{et~al.}  Cover Picture: The Chemical
  Basis of Fungal Bioluminescence. \emph{Angew. Chem. Int. Ed. Engl.}
  \textbf{2015}, \emph{54}, 8001--8001\relax
\mciteBstWouldAddEndPuncttrue
\mciteSetBstMidEndSepPunct{\mcitedefaultmidpunct}
{\mcitedefaultendpunct}{\mcitedefaultseppunct}\relax
\EndOfBibitem
\bibitem[Widder(2010)]{Widder-2010}
Widder,~E.~A. Bioluminescence in the Ocean: Origins of Biological, Chemical,
  and Ecological Diversity. \emph{Science} \textbf{2010}, \emph{328},
  704--708\relax
\mciteBstWouldAddEndPuncttrue
\mciteSetBstMidEndSepPunct{\mcitedefaultmidpunct}
{\mcitedefaultendpunct}{\mcitedefaultseppunct}\relax
\EndOfBibitem
\bibitem[Navizet \latin{et~al.}(2011)Navizet, Liu, Ferr{\'e}, Roca-Sanju{\'a}n,
  and Lindh]{Navizet-2011}
Navizet,~I.; Liu,~Y.-J.; Ferr{\'e},~N.; Roca-Sanju{\'a}n,~D.; Lindh,~R. The
  Chemistry of Bioluminescence: An Analysis of Chemical Functionalities.
  \emph{Chem. Phys. Chem.} \textbf{2011}, \emph{12}, 3064--3076\relax
\mciteBstWouldAddEndPuncttrue
\mciteSetBstMidEndSepPunct{\mcitedefaultmidpunct}
{\mcitedefaultendpunct}{\mcitedefaultseppunct}\relax
\EndOfBibitem
\bibitem[Daunert and Deo(2006)Daunert, and Deo]{Daunert-2006}
Daunert,~S.; Deo,~S.~K. \emph{Photoproteins in Bioanalysis}; Wiley-VCH Verlag
  GmbH \& Co. KGaA, 2006; pp 235--240\relax
\mciteBstWouldAddEndPuncttrue
\mciteSetBstMidEndSepPunct{\mcitedefaultmidpunct}
{\mcitedefaultendpunct}{\mcitedefaultseppunct}\relax
\EndOfBibitem
\bibitem[Ripp \latin{et~al.}(2003)Ripp, Daumer, McKnight, Levine, Garland,
  Simpson, and Sayler]{Ripp-2003}
Ripp,~S.; Daumer,~K.~A.; McKnight,~T.; Levine,~L.~H.; Garland,~J.~L.;
  Simpson,~M.~L.; Sayler,~G.~S. Bioluminescent bioreporter integrated-circuit
  sensing of microbial volatile organic compounds. \emph{J. Ind. Microbiol.
  Biotechnol.} \textbf{2003}, \emph{30}, 636--642\relax
\mciteBstWouldAddEndPuncttrue
\mciteSetBstMidEndSepPunct{\mcitedefaultmidpunct}
{\mcitedefaultendpunct}{\mcitedefaultseppunct}\relax
\EndOfBibitem
\bibitem[De~Vico \latin{et~al.}(2007)De~Vico, Liu, Krogh, and
  Lindh]{DeVico-2007}
De~Vico,~L.; Liu,~Y.-J.; Krogh,~J.~W.; Lindh,~R. Chemiluminescence of
  1,2-Dioxetane. Reaction Mechanism Uncovered. \emph{J. Phys. Chem. A}
  \textbf{2007}, \emph{111}, 8013--8019\relax
\mciteBstWouldAddEndPuncttrue
\mciteSetBstMidEndSepPunct{\mcitedefaultmidpunct}
{\mcitedefaultendpunct}{\mcitedefaultseppunct}\relax
\EndOfBibitem
\bibitem[Farahani \latin{et~al.}(2013)Farahani, Roca-Sanju{\'a}n, Zapata, and
  Lindh]{Farahani-2013}
Farahani,~P.; Roca-Sanju{\'a}n,~D.; Zapata,~F.; Lindh,~R. Revisiting the
  Nonadiabatic Process in 1,2-Dioxetane. \emph{J. Chem. Theory Comput.}
  \textbf{2013}, \emph{9}, 5404--5411\relax
\mciteBstWouldAddEndPuncttrue
\mciteSetBstMidEndSepPunct{\mcitedefaultmidpunct}
{\mcitedefaultendpunct}{\mcitedefaultseppunct}\relax
\EndOfBibitem
\bibitem[Vacher \latin{et~al.}(2017)Vacher, Brakestad, Karlsson,
  Fdez.~Galv{\'a}n, and Lindh]{Vacher-2017-JCTC}
Vacher,~M.; Brakestad,~A.; Karlsson,~H.~O.; Fdez.~Galv{\'a}n,~I.; Lindh,~R.
  Dynamical Insights into the Decomposition of 1,2-Dioxetane. \emph{J. Chem.
  Theory Comput.} \textbf{2017}, \emph{13}, 2448--2457\relax
\mciteBstWouldAddEndPuncttrue
\mciteSetBstMidEndSepPunct{\mcitedefaultmidpunct}
{\mcitedefaultendpunct}{\mcitedefaultseppunct}\relax
\EndOfBibitem
\bibitem[Adam and Baader(1985)Adam, and Baader]{Adam-1985}
Adam,~W.; Baader,~W.~J. Effects of methylation on the thermal stability and
  chemiluminescence properties of 1,2-dioxetanes. \emph{J. Am. Chem. Soc.}
  \textbf{1985}, \emph{107}, 410--416\relax
\mciteBstWouldAddEndPuncttrue
\mciteSetBstMidEndSepPunct{\mcitedefaultmidpunct}
{\mcitedefaultendpunct}{\mcitedefaultseppunct}\relax
\EndOfBibitem
\bibitem[Lourderaj \latin{et~al.}(2008)Lourderaj, Park, and
  Hase]{Lourderaj-2008}
Lourderaj,~U.; Park,~K.; Hase,~W.~L. Classical trajectory simulations of
  post-transition state dynamics. \emph{Int. Rev. Phys. Chem.} \textbf{2008},
  \emph{27}, 361--403\relax
\mciteBstWouldAddEndPuncttrue
\mciteSetBstMidEndSepPunct{\mcitedefaultmidpunct}
{\mcitedefaultendpunct}{\mcitedefaultseppunct}\relax
\EndOfBibitem
\bibitem[Sun \latin{et~al.}(2012)Sun, Park, de~Jong, Lischka, Windus, and
  Hase]{Sun-2012}
Sun,~R.; Park,~K.; de~Jong,~W.~A.; Lischka,~H.; Windus,~T.~L.; Hase,~W.~L.
  Direct dynamics simulation of dioxetane formation and decomposition via the
  singlet $\cdot$O--O--CH2--CH2$\cdot$ biradical: Non-RRKM dynamics. \emph{J.
  Chem. Phys.} \textbf{2012}, \emph{137}, 044305\relax
\mciteBstWouldAddEndPuncttrue
\mciteSetBstMidEndSepPunct{\mcitedefaultmidpunct}
{\mcitedefaultendpunct}{\mcitedefaultseppunct}\relax
\EndOfBibitem
\bibitem[Barbatti \latin{et~al.}(2006)Barbatti, Granucci, Lischka, Persico, and
  Ruckenbauer]{Newton-X}
Barbatti,~M.; Granucci,~G.; Lischka,~H.; Persico,~M.; Ruckenbauer,~M. Newton-X,
  version 0.11b. www.univie.ac.at/newtonx, 2006\relax
\mciteBstWouldAddEndPuncttrue
\mciteSetBstMidEndSepPunct{\mcitedefaultmidpunct}
{\mcitedefaultendpunct}{\mcitedefaultseppunct}\relax
\EndOfBibitem
\bibitem[Tully(1990)]{Tully-1990}
Tully,~J.~C. Molecular dynamics with electronic transitions. \emph{J. Chem.
  Phys.} \textbf{1990}, \emph{93}, 1061--1071\relax
\mciteBstWouldAddEndPuncttrue
\mciteSetBstMidEndSepPunct{\mcitedefaultmidpunct}
{\mcitedefaultendpunct}{\mcitedefaultseppunct}\relax
\EndOfBibitem
\bibitem[Granucci and Persico(2007)Granucci, and Persico]{Granucci-2007}
Granucci,~G.; Persico,~M. Critical appraisal of the fewest switches algorithm
  for surface hopping. \emph{J. Chem. Phys.} \textbf{2007}, \emph{126},
  134114\relax
\mciteBstWouldAddEndPuncttrue
\mciteSetBstMidEndSepPunct{\mcitedefaultmidpunct}
{\mcitedefaultendpunct}{\mcitedefaultseppunct}\relax
\EndOfBibitem
\bibitem[Aquilante \latin{et~al.}(2016)Aquilante, Autschbach, Carlson,
  Chibotaru, Delcey, De~Vico, Fdez.~Galv{\'a}n, Ferr{\'e}, Frutos, Gagliardi,
  Garavelli, Giussani, Hoyer, Li~Manni, Lischka, Ma, Malmqvist, M{\"u}ller,
  Nenov, Olivucci, Pedersen, Peng, Plasser, Pritchard, Reiher, Rivalta,
  Schapiro, Segarra-Mart{\'\i}, Stenrup, Truhlar, Ungur, Valentini, Vancoillie,
  Veryazov, Vysotskiy, Weingart, Zapata, and Lindh]{Molcas-2016}
Aquilante,~F.; Autschbach,~J.; Carlson,~R.~K.; Chibotaru,~L.~F.; Delcey,~M.~G.;
  De~Vico,~L.; Fdez.~Galv{\'a}n,~I.; Ferr{\'e},~N.; Frutos,~L.~M.;
  Gagliardi,~L. \latin{et~al.}  Molcas 8: New capabilities for
  multiconfigurational quantum chemical calculations across the periodic table.
  \emph{J. Comput. Chem.} \textbf{2016}, \emph{37}, 506--541\relax
\mciteBstWouldAddEndPuncttrue
\mciteSetBstMidEndSepPunct{\mcitedefaultmidpunct}
{\mcitedefaultendpunct}{\mcitedefaultseppunct}\relax
\EndOfBibitem
\bibitem[Roos \latin{et~al.}(1980)Roos, Taylor, and Siegbahn]{Roos-1980}
Roos,~B.~O.; Taylor,~P.~R.; Siegbahn,~P.~E. A complete active space SCF method
  (CASSCF) using a density matrix formulated super-CI approach. \emph{Chem.
  Phys.} \textbf{1980}, \emph{48}, 157--173\relax
\mciteBstWouldAddEndPuncttrue
\mciteSetBstMidEndSepPunct{\mcitedefaultmidpunct}
{\mcitedefaultendpunct}{\mcitedefaultseppunct}\relax
\EndOfBibitem
\bibitem[Roos \latin{et~al.}(2004)Roos, Lindh, Malmqvist, Veryazov, and
  Widmark]{Roos-2004}
Roos,~B.~O.; Lindh,~R.; Malmqvist,~P.-{\AA}.; Veryazov,~V.; Widmark,~P.-O. Main
  Group Atoms and Dimers Studied with a New Relativistic ANO Basis Set.
  \emph{J. Phys. Chem. A} \textbf{2004}, \emph{108}, 2851--2858\relax
\mciteBstWouldAddEndPuncttrue
\mciteSetBstMidEndSepPunct{\mcitedefaultmidpunct}
{\mcitedefaultendpunct}{\mcitedefaultseppunct}\relax
\EndOfBibitem
\bibitem[Aquilante \latin{et~al.}(2009)Aquilante, Gagliardi, Pedersen, and
  Lindh]{Aquilante-2009}
Aquilante,~F.; Gagliardi,~L.; Pedersen,~T.~B.; Lindh,~R. Atomic Cholesky
  decompositions: A route to unbiased auxiliary basis sets for density fitting
  approximation with tunable accuracy and efficiency. \emph{J. Chem. Phys.}
  \textbf{2009}, \emph{130}, 154107\relax
\mciteBstWouldAddEndPuncttrue
\mciteSetBstMidEndSepPunct{\mcitedefaultmidpunct}
{\mcitedefaultendpunct}{\mcitedefaultseppunct}\relax
\EndOfBibitem
\bibitem[Bastos \latin{et~al.}(2013)Bastos, da~Silva, and Baader]{Bastos-2013}
Bastos,~E.~L.; da~Silva,~S.~M.; Baader,~W.~J. Solvent Cage Effects: Basis of a
  General Mechanism for Efficient Chemiluminescence. \emph{J. Org. Chem.}
  \textbf{2013}, \emph{78}, 4432--4439\relax
\mciteBstWouldAddEndPuncttrue
\mciteSetBstMidEndSepPunct{\mcitedefaultmidpunct}
{\mcitedefaultendpunct}{\mcitedefaultseppunct}\relax
\EndOfBibitem
\end{mcitethebibliography}

\end{document}